\documentclass[12pt,reqno]{amsart}
\usepackage{amsmath}
\usepackage{amssymb}
\usepackage{verbatim}
\usepackage{mathrsfs}
\usepackage{graphicx} 
\usepackage[usenames,dvipsnames]{pstricks}
\usepackage{pst-plot,pstricks-add}

\usepackage[top=0.6in,bottom=0.6in,left=0.7in,right=0.7in]{geometry}

\newcommand{\C}{\mathbb{C}}    
\newcommand{\N}{\mathbb{N}}    
\newcommand{\R}{\mathbb{R}}    
\newcommand{\Z}{\mathbb{Z}}    
\newcommand{\dR}{\mathbb{R}^d}

\newcommand{\dZ}{\mathbb{Z}^d}
\newcommand{\dlp}[1]{l_{#1}(\mathbb{Z}^d)}
\newcommand{\td}{\boldsymbol{\delta}}  
\newcommand{\bp}{ \begin{proof} }
\newcommand{\ep}{\hfill  \end{proof} }
\newcommand{\be}{ \begin{equation} }
\newcommand{\ee}{ \end{equation} }
\newcommand{\dLp}[1]{L_{#1}(\mathbb{R}^d)}
\newcommand{\prm}{P}           
\newcommand{\wh}{\widehat}
\renewcommand{\le}{\leqslant}

\newcommand{\bs}{\backslash}
\newcommand{\ol}{\overline}
\newcommand{\la}{\langle}
\newcommand{\ra}{\rangle}
\newcommand{\tp}{\mathsf{T}}  
\newcommand{\setsp}{\;:\;}     

\newtheorem{lemma}{Lemma}
\newtheorem{theorem}[lemma]{Theorem}
\newtheorem{example}{Example}

\begin{document}

\title[Directional Box Spline Tight Framelets]{Directional Compactly supported Box Spline Tight Framelets with Simple Structure}

\author{Bin Han}

\address{Department of Mathematical and Statistical Sciences,
University of Alberta, Edmonton,\quad Alberta, Canada T6G 2G1.
\quad {\tt bhan@ualberta.ca},\quad {\tt tao8@ualberta.ca}
}

\author{Tao Li}


\author{Xiaosheng Zhuang}
\address{Department of Mathematics, City University of Hong Kong, Tat Chee Avenue, Kowloon Tong, Hong Kong.
\quad {\tt xzhuang7@cityu.edu.hk}.}

\thanks{
Research of B. Han is supported in part by NSERC Canada under Grant RGP 228051. Research of T. Li is supported by a Mitacs global link project. Research of X. Zhuang is supported in part by the Research Grants Council of Hong Kong (Project No. CityU 11304414).
}

\makeatletter \@addtoreset{equation}{section} \makeatother
\begin{abstract}
To effectively capture singularities in high-dimensional data and functions, multivariate compactly supported tight framelets, having directionality and derived from refinable box splines, are of particular interest in both theory and applications. The $d$-dimensional Haar refinable function $\chi_{[0,1]^d}$ is a simple example of refinable box splines. For every dimension $d\in \N$, in this paper we construct a directional compactly supported $d$-dimensional Haar tight framelet such that all its high-pass filters in its underlying tight framelet filter bank have only two nonzero coefficients with opposite signs and they exhibit totally $(3^d-1)/2$ directions in dimension $d$.
Furthermore, applying the projection method to such directional Haar tight framelets, from every refinable box spline in every dimension, we construct a directional compactly supported box spline tight framelet with simple structure such that all the high-pass filters in its underlying tight framelet filter bank have only two nonzero coefficients with opposite signs. Moreover, such compactly supported box spline tight framelets can achieve arbitrarily high numbers of directions by using refinable box splines with increasing supports.
\end{abstract}


 \maketitle

\pagenumbering{arabic}


To capture singularities in many high-dimensional data such as images/videos, directional representations are of great importance in both theory and applications, for example, see curvelets and shearlets in \cite{CD04,GKL06} and tensor product complex tight framelets in \cite{han13,hz14}.
On the other hand, (refinable) box splines are widely used in both approximation theory and wavelet analysis. Motivated by the interesting example of a two-dimensional directional Haar tight framelet constructed in \cite{lcshi16} which has impressive performance in parallel magnetic resonance imaging (pMRI), in this paper we construct compactly supported tight framelets with directionality and very simple structures from the Haar refinable functions and all refinable box splines in all dimensions. All the high-pass filters in such directional tight framelets have only two nonzero coefficients with oppositive signs. Consequently, all of them naturally exhibit directionality and their associated fast framelet transforms can be efficiently implemented through simple difference operations.

Let us first recall some definitions and notation.
By $\dlp{0}$ we denote the set of all finitely supported sequences/filters $a=\{a(k)\}_{k\in \Z}: \dZ\rightarrow \C$ on $\dZ$. For a filter $a\in \dlp{0}$, its Fourier series is defined to be $\wh{a}(\xi):=\sum_{k\in \dZ} a(k)e^{-ik\cdot\xi}$ for $\xi\in \dR$, which is a $2\pi\dZ$-periodic trigonometric polynomial. In particular, by $\td$ we denote \emph{the Dirac sequence} such that $\td(0)=1$ and $\td(k)=0$ for all $\dZ\bs\{0\}$. For $\gamma\in \dZ$, we also use the notation $\td_\gamma$ to stand for the sequence $\td(\cdot-\gamma)$, i.e., $\td_\gamma(\gamma)=1$ and $\td_\gamma(k)=0$ for all $k\in \dZ\bs \{\gamma\}$.
Note that $\wh{\td_\gamma}(\xi)=e^{-i\gamma\cdot \xi}$. For filters $a,b_1,\ldots,b_s\in \dlp{0}$,
we say that a filter bank $\{a;b_1,\ldots,b_s\}$ is \emph{a ($d$-dimensional dyadic) tight framelet filter bank} if
\be \label{tffb}
\wh{a}(\xi)\ol{\wh{a}(\xi+\pi \omega)}+
\sum_{\ell=1}^s \wh{b_\ell}(\xi)\ol{\wh{b_\ell}(\xi+\pi \omega)}=\td(\omega),\qquad \forall\, \xi\in \dR, \omega\in \{0,1\}^d.
\ee
In the spatial domain, the equations in \eqref{tffb} can be equivalently rewritten as
\be \label{tffb:2}
\sum_{k\in \dZ} a(\gamma+2k)\ol{a(n+\gamma+2k)}+
\sum_{\ell=1}^s \sum_{k\in \dZ}
b_\ell(\gamma+2k)\ol{b_\ell(n+\gamma+2k)}=
2^{-d}\td(n),\qquad \forall\, \gamma\in \{0,1\}^d
\ee
for all $n\in \dZ$.
Let $\phi,\psi_1,\ldots,\psi_s\in \dLp{2}$. We say that $\{\phi;\psi_1,\ldots,\psi_s\}$ is \emph{a (nonhomogeneous) tight framelet in $\dLp{2}$} if
\be \label{tf}
\|f\|_{\dLp{2}}^2=\sum_{k\in \dZ} |\la f,\phi(\cdot-k)\ra|^2+
\sum_{j=0}^\infty \sum_{\ell=1}^s \sum_{k\in \dZ} |\la f, 2^{jd/2}\psi_\ell(2^j\cdot-k)\ra|^2,\qquad \forall \, f\in \dLp{2}.
\ee
Let $\{a;b_1,\ldots,b_s\}$ be a ($d$-dimensional dyadic) tight framelet filter bank with all filters $a,b_1,\ldots,b_s\in \dlp{0}$ such that $\wh{a}(0)=\sum_{k\in \dZ} a(k)=1$.
Then we can define compactly supported tempered distributions $\phi$ and $\psi_1,\ldots,\psi_s$ on $\dR$ through
\be \label{phi:psi}
\wh{\phi}(\xi):=\prod_{j=1}^\infty \wh{a}(2^{-j}\xi),\qquad \xi\in \dR\quad \mbox{and}\quad
\wh{\psi_\ell}(\xi)=\wh{b_\ell}(\xi/2)\wh{\phi}(\xi/2),\qquad \xi\in \dR, \ell=1,\ldots,s.
\ee
Then we must have $\phi,\psi_1,\ldots,\psi_s\in \dLp{2}$ and $\{\phi;\psi_1,\ldots,\psi_s\}$ is a ($d$-dimensional dyadic) tight framelet in $\dLp{2}$ (see \cite[Lemma~2.1]{han03},
\cite[Corollary~12 and Theorem~17]{han12} and \cite{dhrs03,rs97}).
See \cite{chs02,dgm86,dhrs03,han12,han03,han97,ls06,rs97} and many references therein
for extensive investigation on tight framelets derived from refinable functions.
The tempered distribution $\phi$ in \eqref{phi:psi} is called \emph{a refinable function} satisfying the refinement equation
$\wh{\phi}(\xi)=\wh{a}(\xi/2)\wh{\phi}(\xi/2)$ for $\xi\in \dR$ with the refinement filter $a$.

This paper is motivated by the interesting paper \cite{lcshi16}, where a two-dimensional directional Haar tight framelet has been constructed and applied with impressive performance to pMRI. Applying finite linear combinations to the standard tensor product two-dimensional Haar wavelet,
in \cite[(3.5)]{lcshi16}, the authors constructed a two-dimensional directional tight framelet filter bank $\{a^H;b_1,\ldots,b_6\}$, where $a^H$ is the two-dimensional Haar low-pass filter and
\begin{align*}
&b_1(k)=\begin{cases}
\frac{1}{4} &\text{if $k=(0,0)$},\\
-\frac{1}{4} &\text{if $k=(1,1)$},\\
0 &\text{otherwise},
\end{cases}
&&b_2(k)=\begin{cases}
\frac{1}{4} &\text{if $k=(1,0)$},\\
-\frac{1}{4} &\text{if $k=(0,1)$},\\
0 &\text{otherwise},
\end{cases}
&&b_3(k)=\begin{cases}
\frac{1}{4} &\text{if $k=(0,0)$},\\
-\frac{1}{4} &\text{if $k=(0,1)$},\\
0 &\text{otherwise},
\end{cases}\\
&b_4(k)=\begin{cases}
\frac{1}{4} &\text{if $k=(0,0)$},\\
-\frac{1}{4} &\text{if $k=(1,0)$},\\
0 &\text{otherwise},
\end{cases}
&&b_5(k)=\begin{cases}
\frac{1}{4} &\text{if $k=(1,0)$},\\
-\frac{1}{4} &\text{if $k=(1,1)$},\\
0 &\text{otherwise},
\end{cases}
&&b_6(k)=\begin{cases}
\frac{1}{4} &\text{if $k=(0,1)$},\\
-\frac{1}{4} &\text{if $k=(1,1)$},\\
0 &\text{otherwise}.
\end{cases}
\end{align*}
Here the $d$-dimensional Haar low-pass filter is given by
\be \label{a:haar}
a^H(k)=\begin{cases}
2^{-d} &\text{if $k\in \{0,1\}^d$},\\
0 &\text{otherwise}.
\end{cases}
\ee
Since each high-pass filter has only two nonzero coefficients with opposite signs, it naturally has directionality and very simple structures.
To deal with problems such as video processing in dimensions higher than two, it is very natural to ask

\smallskip

\noindent \textbf{Q1.} Is it possible to construct a directional Haar tight framelet for every dimension such that each high-pass filter has only two nonzero coefficients with opposite signs?

\smallskip

A similar construction  method/argument as in \cite{lcshi16} will quickly run into difficulty, since there are so many possible linear combinations even at the dimension three.
Fortunately, adopting a geometric viewpoint,
we can positively answer the question \textbf{Q1} completely as follows:

\begin{theorem}\label{thm:haar}
Let $a^H$ be the $d$-dimensional Haar low-pass filter in \eqref{a:haar}. Define the high-pass filters $b_1,\ldots,b_s$ with $s:=\binom{2^d}{2}=2^{d-1}(2^d-1)$ in the following way: $2^{-d}(\td_{\gamma_1}-\td_{\gamma_2})$ for all
undirected edges with endpoints $\gamma_1,\gamma_2\in \{0,1\}^d$ and $\gamma_1\ne \gamma_2$. Then $\{a^H; b_1,\ldots,b_s\}$ is a tight framelet filter bank such that all the high-pass filters $b_1,\ldots,b_s$ have directionality and exhibit totally $\frac{1}{2}(3^d-1)$ directions in dimension $d$. Define functions $\phi$ and $\psi_1,\ldots,\psi_s$ as in \eqref{phi:psi}. Then $\{\phi;\psi_1,\ldots,\psi_s\}$ is a $d$-dimensional directional compactly supported Haar tight framelet in $\dLp{2}$.
\end{theorem}

\bp To prove that $\{a^H; b_1,\ldots,b_s\}$ is a tight framelet filter bank, we have to check the conditions in \eqref{tffb:2} for all $\gamma\in \{0,1\}^d$ and $n\in \dZ$. Since all the filters are supported inside $\{0,1\}^d$, it is trivial to observe that all the filters $a^H,b_1,\ldots,b_s$ vanish at the position $\gamma+2k$ for all $k\in \dZ\bs\{0\}$. Therefore, the equations \eqref{tffb:2} become
\be \label{tffb:haar}
a^H(\gamma)\ol{a^H(n+\gamma)}+
\sum_{\ell=1}^s
b_\ell(\gamma)\ol{b_\ell(n+\gamma)}=
2^{-d}\td(n),\qquad \gamma\in \{0,1\}^d, n\in \dZ.
\ee

Case 1: $n=0$. Then $|a^H(\gamma)|^2=2^{-2d}$. By the definition of the high-pass filters, there are totally $2^d-1$ filters whose supports contain the point $\gamma$. Consequently, $\sum_{\ell=1}^s |b_\ell(\gamma)|^2=2^{-2d}(2^d-1)$. Thus, we have $|a^H(\gamma)|^2+\sum_{\ell=1}^s |b_\ell(\gamma)|^2= 2^{-2d}+2^{-2d}(2^d-1)=2^{-d}$
which proves \eqref{tffb:haar} with $n=0$.

Case 2: $n\ne 0$ and $n+\gamma \not \in \{0,1\}^d$. For this case, since all the filters
are supported inside $\{0,1\}^d$,
we trivially have $a^H(n+\gamma)=0$ and $b_\ell(n+\gamma)=0$ for all $\ell=1,\ldots,s$. Hence, \eqref{tffb:haar} is trivially true for $n\ne 0$ and $n+\gamma \not \in \{0,1\}^d$.

Case 3: $n\ne 0$ and $n+\gamma \in \{0,1\}^d$. Then $n$ and $n+\gamma$ are two distinct points in $\{0,1\}^d$. By the definition of the high-pass filters, there exists exactly one integer $j$ with $1\le j\le s$ such that
$b_j(\gamma)\ol{b_j(n+\gamma)}=-2^{-2d}$ and
$b_\ell(\gamma)\ol{b_\ell(n+\gamma)}=0$ for all $\ell \in \{1,\ldots,s\}\bs \{j\}$. Noting that $a^H(\gamma)=a^H(n+\gamma)=2^{-d}$, we conclude
\[
a^H(\gamma)\ol{a^H(n+\gamma)}+
\sum_{\ell=1}^s
b_\ell(\gamma)\ol{b_\ell(n+\gamma)}=
a^H(\gamma)\ol{a^H(n+\gamma)}+
b_j(\gamma)\ol{b_j(n+\gamma)}
=2^{-2d}-2^{-2d}=0,
\]
which proves \eqref{tffb:haar} for $n\ne 0$ and $n+\gamma\in \{0,1\}$.

Therefore, $\{a^H; b_1,\ldots,b_s\}$ is a tight framelet filter bank. Since each high-pass filter has only two nonzero coefficients with opposite signs, all the high-pass filters $b_1,\ldots,b_s$ trivially have directionality.
We now count the total number of directions of all the high-pass filters.
Note that the direction of a high-pass filter $2^{-d}(\td_{\gamma_1}-\td_{\gamma_2})$ can be represented by the vector $v=\gamma_1-\gamma_2$ or $v=\gamma_2-\gamma_1$. Such a direction vector $v$ is unique if we additionally require that the first nonzero entry of $v$ should be positive.
Note that $v\in \{-1,0,1\}^d\bs\{0\}$ with each entry of $v$ belonging to $\{-1,0,1\}$.
Let $S$ be the set of all the nonzero vectors $v\in \{-1,0,1\}^d$ such that the first nonzero entry of $v$ is positive (i.e., $1$).  Hence, any direction vector $v$ of a high-pass filter belongs to $S$.
Conversely, for every vector $v\in S$, we can uniquely write $v=\gamma_1-\gamma_2$ with $\gamma_1,\gamma_2\in \{0,1\}^d$ and $\gamma_1+\gamma_2\in \{0,1\}^d$ by separating the positive and negative entries of $v$. Therefore, the vector $v$ represents the direction of the high-pass filter $2^{-d}(\td_{\gamma_1}- \td_{\gamma_2})$. Hence, the total number of directions of all the high-pass filters is equal to the cardinality of the set $S$.
Consider the subset $S_j$ whose elements are in $S$ with the first nonzero entry at the position $j$ for $j=1,\ldots, d$. Clearly, the cardinality of $S_j$ is $3^{d-j}$. Since $S$ is the disjoint union of $S_1,\ldots, S_d$, we conclude that the cardinality of $S$ is
$3^{d-1}+3^{d-2}+\cdots +3^{d-d}=(3^d-1)/2$.
\ep

The tight framelet filter bank in Theorem~\ref{thm:haar}
with $d=1$ is just the standard Haar orthogonal wavelet filter bank and the case $d=2$ recovers the directional Haar tight framelet filter bank in \cite{lcshi16}. We now provide an alternative algebraic proof to Theorem~\ref{thm:haar} from the viewpoint of discrete framelet transforms. Because all the filters in Theorem~\ref{thm:haar} are supported inside $\{0,1\}^d$, the $d$-dimensional discrete framelet transform using the filter bank $\{a;b_1,\ldots,b_s\}$ in Theorem~\ref{thm:haar} can be simply implemented by applying the discrete framelet transform acting on data supported on each disjoint $\{0,1\}^d+2k$, $k\in \dZ$, where $\{0,1\}^d$ is the set of all vertices of the unit cube $[0,1]^d$. For simplicity, we list the vertices as $\{v_1,\ldots,v_{2^d}\}=\{0,1\}^d$ and assume that the data value at the point $v_j$ is $x_j\in \R$.
Then all the high-pass filters in Theorem~\ref{thm:haar} are given by $\pm 2^{-d} (\td_{v_j}-\td_{v_k})$ with $1\le j<k\le 2^d$. The framelet coefficient produced by this high-pass filter is simply $\pm 2^{d/2} 2^{-d} (x_j-x_k)$ (see \cite{han13} for discrete framelet transforms). The coefficient produced by the Haar low-pass filter $a^H$ in Theorem~\ref{thm:haar} is $2^{d/2} 2^{-d} (x_1+\cdots+x_{2^d})$.
Hence, the total squared energy of all the framelet coefficients is
$\left(2^{d/2} 2^{-d} (x_1+\cdots+x_{2^d})\right)^2
+
\sum_{1\le j<k\le 2^d} \left(\pm 2^{d/2} 2^{-d} (x_j-x_k)\right)^2$.
Noting that $(x_1+\cdots+x_{2^d})^2=x_1^2+\cdots+x_{2^d}^2+
\sum_{1\le j<k\le 2^d} 2x_j x_k$ and $(x_j-x_k)^2=(x_j^2+x_k^2)-2x_jx_k$, we conclude that the total squared energy of all the framelet coefficients is
\begin{align*}
\Big (2^{d/2} 2^{-d} (x_1+\cdots+x_{2^d})\Big)^2
&+
\sum_{1\le j<k\le 2^d} \Big(\pm 2^{d/2} 2^{-d} (x_j-x_k)\Big)^2\\
&=2^{-d} (x_1^2+\cdots+x_{2^d}^2)+
2^{-d} \sum_{1\le j<k\le 2^d} (x_j^2+x_k^2)
=x_1^2+\cdots+x_{2^d}^2,
\end{align*}
which proves the energy preservation property of the discrete framelet transform.
By \cite[Theorem~2.4]{han13}, the filter bank $\{a;b_1,\ldots,b_s\}$ in Theorem~\ref{thm:haar} must be a $d$-dimensional tight framelet filter bank.

One possible shortcoming of Theorem~\ref{thm:haar} is that all the Haar refinable functions $\chi_{[0,1]^d}$ are discontinuous and often introduce unpleasant block effects in data/image processing. Therefore, smooth refinable functions and directional smooth tight framelets are often preferred in applications. Refinable box splines can be made arbitrarily smooth and are widely used in approximation theory and wavelet analysis. This naturally leads us to ask

\smallskip

\textbf{Q2.} Can we construct directional compactly supported tight framelets in $\dLp{2}$ with simple structure from every refinable box spline in every dimension?

\smallskip

Using Theorem~\ref{thm:haar} and the projection method, we positively answer the question \textbf{Q2} painlessly. To do so, let us recall the definition of box splines which are closely linked to the projection method.
Let $\prm$ be a $d\times n$ real-valued matrix of rank $d$ with $d\le n$.
A box spline $M_\prm$ with the $d\times n$ direction matrix $\prm$ is defined to be
\be \label{boxspline}
\wh{M_\prm}(\xi):=\prod_{k\in \prm} \frac{1-e^{-ik\cdot \xi}}{ik\cdot \xi},\qquad \xi\in \dR,
\ee
where $k\in \prm$ means that $k$ is a column vector of $\prm$ and $k$ goes through all the columns of $\prm$ once and only once. A box spline can be also defined through the projection method (\cite{han06,han14}).
For an integrable function $f\in L_1(\R^n)$, we can define the projected function $\prm f$ on $\dR$ by
\be \label{F:prm}
\wh{\prm f}(\xi):=\wh{f}(\prm^\tp \xi), \qquad \xi\in \dR.
\ee
Since $\wh{f}$ is continuous on $\R^n$, the function $\wh{\prm f}$ is a well-defined continuous function on $\dR$. In the spatial domain, the definition of the $d$-dimensional projected function $\prm f$ in \eqref{F:prm} can be equivalently expressed as
\be \label{prm}
[\prm f](x)=\frac{1}{\sqrt{\det(\prm \prm^\tp)}} \int_{\prm^{-1} x} f d S,\qquad x\in \dR,
\ee
where $S$ is the surface element on the superplane $\prm^{-1}x:=\{y\in \R^n \setsp \prm y=x\}$. In fact, for $f\in L_1(\R^n)$, the projected function $\prm f\in \dLp{1}$ (see \cite{han14,han06}).
Note that $\wh{\chi_{[0,1]^n}}(\xi)=\prod_{k\in \{0,1\}^n} \frac{1-e^{-ik\cdot \xi}}{ik\cdot \xi}$ and
\[
\wh{\prm \chi_{[0,1]^n}}(\xi)=
\wh{\chi_{[0,1]^n}}(\prm^\tp \xi)=\prod_{k\in \{0,1\}^n} \frac{1-e^{-ik\cdot (\prm^\tp \xi)}}{ik\cdot (\prm^\tp \xi)}
=\prod_{k\in \{0,1\}^n} \frac{1-e^{-i(\prm k)\cdot \xi}}{i(\prm k)\cdot \xi}=\wh{M_\prm}(\xi).
\]
Hence, the box spline $M_\prm$ is nothing else but the projected function $\prm \chi_{[0,1]^n}$ of the $n$-dimensional Haar function along the direction matrix $\prm$. See the book \cite{bhr93} for extensive study on box splines.

The projection method can be also applied to filters on $\Z^n$ provided that $\prm$ is a $d\times n$ integer matrix.
For an $n$-dimensional filter $a\in l_0(\Z^n)$,
the projected filter $\prm a\in \dlp{0}$ is defined by
\be \label{F:prm:filter}
\wh{\prm a}(\xi):=\wh{a}(\prm^\tp \xi), \qquad \xi\in \dR,\quad \mbox{or equivalently},\quad
[\prm a](j)=\sum_{k\in \prm^{-1}j} a(k), \qquad j\in \dZ,
\ee
where $\prm^{-1} j:=\{ k\in \Z^n \setsp \prm k=j\}$.
Because $\wh{a}$ is a $2\pi \Z^n$-periodic trigonometric polynomial and $\prm$ is an integer matrix, $\wh{\prm a}$ is a well-defined $2\pi\Z^d$-periodic trigonometric polynomial.
%
%
If $a^H$ is the $n$-dimensional Haar low-pass filter in \eqref{a:haar}, then we define $a_\prm:=\prm a^H$ to be the box spline refinement filter/mask for the box spline $M_\prm$ with a $d\times n$ direction matrix.
For an integer projection matrix $\prm$,
the box spline function $M_\prm$ in \eqref{boxspline} is refinable: $\wh{M_\prm} (2\xi)=\wh{a_\prm}(\xi) \wh{M_\prm}(\xi)$, since $\prm \chi_{[0,1]^n}=M_\prm$ and the $n$-dimensional Haar function $\chi_{[0,1]^n}$ is refinable: $\wh{\chi_{[0,1]^n}}(2\xi)=\wh{a^H}(\xi)\wh{\chi_{[0,1]^n}}(\xi)$.
Various compactly supported tight framelets have been constructed from refinable box splines with integer direction matrices in the literature, e.g., see
Ron and Shen~\cite{rs98}, Lai and St\"ockler~\cite{ls06}, Han~\cite{han14,han06}, and Fan, Ji and Shen \cite{fjs16}.

Suppose that $\prm $ is a $d\times n$ integer matrix of rank $d$ with $d\le n$ satisfying
\be \label{cond:prm}
\prm^\tp (\dZ\bs [2\dZ])\subseteq \Z^n \bs [2\Z^n].
\ee
For every tight framelet $\{\phi;\psi_1,\ldots,\psi_s\}$ in $L_2(\R^n)$ regardless whether $\{\phi;\psi_1,\ldots,\psi_s\}$ has an associated underlying filter bank or not, it is known in
\cite[Theorem~4]{han06} and
\cite[Theorem~2.3 and Corollary~5.3]{han14}
that $\{\prm \phi; \prm \psi_1,\ldots, \prm \psi_s\}$ must be a tight framelet in $\dLp{2}$.
Similarly, for every $n$-dimensional tight framelet filter bank $\{a; b_1,\ldots,b_s\}$, then
$\{\prm a; \prm b_1,\ldots, \prm b_s\}$ must be a $d$-dimensional tight framelet filter bank. For completeness, let us recall the argument from \cite{han14,han06} here. By definition, $\{a;b_1,\ldots,b_s\}$ is an $n$-dimensional tight framelet filter bank if and only if
\be \label{tffb:n}
\wh{a}(\zeta)\ol{\wh{a}(\zeta+\pi \beta)}+\sum_{\ell=1}^s \wh{b_\ell}(\zeta)\ol{\wh{b_\ell}(\zeta+\pi \beta)}=\td(\beta),\qquad \forall\, \zeta\in \R^n, \beta\in \{0,1\}^n.
\ee
The condition in \eqref{cond:prm} is equivalent to saying that $\prm^\tp \omega\not\in 2\Z^n$ for all $\omega\in \{0,1\}^d\bs\{0\}$, Consequently, for every $\omega\in \{0,1\}^d\bs\{0\}$, we must have
$\prm^\tp \omega \in \Z^n\bs [2\Z^n]$. Therefore, it is trivial to deduce from \eqref{tffb:n} with $\zeta=\prm^\tp \xi$ and $\beta=\prm^\tp \omega$ that
\[
\wh{a}(\prm^\tp \xi)\ol{\wh{a}(\prm^\tp\xi +\pi \prm^\tp \omega)}+\sum_{\ell=1}^s \wh{b_\ell}(\prm^\tp \xi)\ol{\wh{b_\ell}(\prm^\tp\xi+\pi \prm^\tp \omega)}=\td(\omega),\qquad \forall\, \xi\in \R^d, \omega\in \{0,1\}^d.
\]
By the definition of the projected filters in \eqref{F:prm:filter}, the above equations simply become
\[
\wh{\prm a}(\xi)\ol{\wh{\prm a}(\xi +\pi \omega)}+\sum_{\ell=1}^s \wh{\prm b_\ell}(\xi)\ol{\wh{\prm b_\ell}(\xi+\pi \omega)}=\td(\omega),\qquad \forall\, \xi\in \R^d, \omega\in \{0,1\}^d.
\]
That is, $\{\prm a; \prm b_1,\ldots, \prm b_s\}$ must be a $d$-dimensional tight framelet filter bank.

The condition in \eqref{cond:prm} is equivalent to saying (\cite[Theorem~2.5]{han14}) that the filter $a_\prm$ has the sum rules of order at least one,
i.e., $\sum_{k\in \dZ} a_\prm(\gamma+2k)=2^{-d}$ for all $\gamma\in \{0,1\}^d$.
If the condition in \eqref{cond:prm} fails, then the box spline filter $a_\prm$ does not have any sum rules and therefore, no tight framelets can be ever derived from the box spline $M_\prm$, see \cite[Theorem~2.5]{han14} for more detail.
The projection method is originally introduced in Han~\cite{han99} to study optimal smoothness of interpolating refinable functions and has been further developed in \cite{han03,han06} and other papers to study refinable functions, biorthogonal wavelets, and framelets. See Han \cite{han06,han14} for some applications of the projection method in wavelet analysis.

Note that if a high-pass filter $b$ has only two nonzero coefficients with opposite signs, then either $\prm b=0$ or $\prm b$ has only two nonzero coefficients with opposite signs.
Applying the projection method to the Haar tight framelets in Theorem~\ref{thm:haar},
we have the following result positively answering \textbf{Q2}.

\begin{theorem}\label{thm:boxspline}
Let $\prm $ be a $d\times n$ integer matrix of rank $d$ with $d\le n$ such that \eqref{cond:prm} holds.
Let $\{a^H;b_1,\ldots,b_s\}$ with $s:=\binom{2^n}{2}=2^{n-1}(2^n-1)$ be the $n$-dimensional Haar tight framelet filter bank constructed in Theorem~\ref{thm:haar}. Then
$\{\prm a^H; \prm b_1,\ldots,\prm b_s\}$ is a $d$-dimensional tight framelet filter bank with $\prm a^H$ being the box spline refinement filter $a_\prm$ such that all the high-pass filters have only two nonzero coefficients with opposite signs.
Define $\phi$ and $\psi_1,\ldots,\psi_s$ as in \eqref{phi:psi} with $a$ and $b_1,\ldots,b_s$ being replaced by $\prm a^H$ and $\prm b_1,\ldots,\prm b_s$, respectively. Then $\{\phi;\psi_1,\ldots,\psi_s\}$ is a $d$-dimensional directional compactly supported tight framelet in $\dLp{2}$ with $\phi$ being the box spline $M_\prm$ in \eqref{boxspline} having the direction matrix $\prm$.
\end{theorem}

Let $\{a;b_1,\ldots,b_s\}$ be a $d$-dimensional tight framelet filter bank.
If $b_1=c_1b$ and $b_2=c_2 b$ for some constants $c_1, c_2\in \C$ and $b\in \dlp{0}$, then it is trivial that $\{a; \sqrt{|c_1|^2+|c_2|^2} b, b_3,\ldots, b_s\}$ is a tight framelet filter bank. That is, we
can combine high-pass filters which are almost
the same up to a multiplicative constant.
Hence, the number of high-pass filters in Theorem~\ref{thm:boxspline} can be reduced.
Now we provide a geometric construction for the box spline tight framelet filter bank in Theorem~\ref{thm:boxspline} but with similar filters combined to reduce the number of filters.
First we calculate the support $\mbox{supp}(a_\prm)$ of the box spline filter $a_\prm$, which must be the set $\prm \{0,1\}^n \subseteq \dZ$.
Now the set $\{0,1\}^n$ of the vertices of the unit cube $[0,1]^n$ can be written as a disjoint union of the subsets $\prm^{-1} k, k\in \mbox{supp}(a_\prm)$. Then all the high-pass filters are constructed in the following way:
For every pair of two distinct points $\gamma_1, \gamma_2\in \mbox{supp}(a_\prm)$, construct the high-pass filter $2^{-n}\sqrt{(\#\prm^{-1}\gamma_1)(\#\prm^{-1}\gamma_2)}
(\td_{\gamma_1}-\td_{\gamma_2})$, where $\#S$ is the cardinality of a set $S$. Clearly, there are totally $\binom{m}{2}$ with $m:=\# \mbox{supp}(a_\prm)$ number of high-pass filters.
We now provide two examples to illustrate this construction.

\begin{figure}[ht]
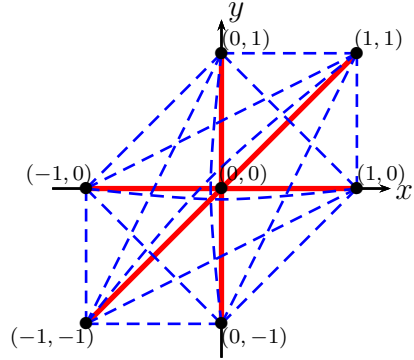

\psset{yunit=1.8cm,xunit=1.8cm}
\pspicture[](-1.2,-1.2)(1.25,1.25)
\psaxes[labels=none, ticks=none]{->}(0,0)(-1.25,-1.25)(1.25,1.25)
\rput[b](1.35,-0.08){$x$}
\rput[l](0.05,1.3){$y$}
\psline[linecolor=red,linewidth=2pt](0,0)(-1,0)
\psline[linecolor=red,linewidth=2pt](0,0)(0,-1)
\psline[linecolor=red,linewidth=2pt](0,0)(1,0)
\psline[linecolor=red,linewidth=2pt](0,0)(0,1)
\psline[linecolor=red,linewidth=2pt](0,0)(1,1)
\psline[linecolor=red,linewidth=2pt](0,0)(-1,-1)
%
\psline[linecolor=blue,linewidth=1pt,linestyle=dashed](0,1)(1,1)
\psline[linecolor=blue,linewidth=1pt,linestyle=dashed](1,1)(1,0)
\psline[linecolor=blue,linewidth=1pt,linestyle=dashed](1,0)(0,-1)
\psline[linecolor=blue,linewidth=1pt,linestyle=dashed](0,-1)(-1,-1)
\psline[linecolor=blue,linewidth=1pt,linestyle=dashed](-1,-1)(-1,0)
\psline[linecolor=blue,linewidth=1pt,linestyle=dashed](-1,0)(0,1)
%
\psline[linecolor=blue,linewidth=1pt,linestyle=dashed](0,1)(1,0)
\psline[linecolor=blue,linewidth=1pt,linestyle=dashed](-1,0)(0,-1)
\psline[linecolor=blue,linewidth=1pt,linestyle=dashed](0,1)(-1,-1)
\psline[linecolor=blue,linewidth=1pt,linestyle=dashed](0,-1)(1,1)
\psline[linecolor=blue,linewidth=1pt,linestyle=dashed](-1,0)(1,1)
\psline[linecolor=blue,linewidth=1pt,linestyle=dashed](-1,-1)(1,0)
%
\pscurve[linecolor=blue,linewidth=1pt,linestyle=dashed](-1,-1)(-0.08,0.08)(1,1)
\pscurve[linecolor=blue,linewidth=1pt,linestyle=dashed](-1,0)(0,-0.08)(1,0)
\pscurve[linecolor=blue,linewidth=1pt,linestyle=dashed](0,-1)(-0.08,0)(0,1)
\rput(0,0){\black $\bullet$}\rput(0.16,0.1){\tiny $(0,0)$}
\rput(1,0){\black $\bullet$}\rput(1.18,0.1){\tiny $(1,0)$}
\rput(0,1){\black $\bullet$}\rput(0.18,1.1){\tiny $(0,1)$}
\rput(1,1){\black $\bullet$}\rput(1.16,1.1){\tiny $(1,1)$}
\rput(-1,-1){\black $\bullet$}\rput(-1.25,-1.1){\tiny $(-1,-1)$}
\rput(0,-1){\black $\bullet$}\rput(0.24,-1.1){\tiny $(0,-1)$}
\rput(-1,0){\black $\bullet$}\rput(-1.2,0.1){\tiny $(-1,0)$}
\endpspicture
\caption{Each edge connecting every two vertices indicates a high-pass filter with coefficients of the same weight but opposite signs at its two endpoints.
Weight for each dashed blue edge (total $15$) is $\frac18$, while weight for each solid red edge (total $6$) is $\frac{\sqrt{2}}{8}$. The total number of all the high-pass filters is $21$ with $6$ solid red edges and $15$ dashed blue edges. The total number of directions/slopes for the $21$ high-pass filters is $6$ with angles $0^\circ$ ($5$ edges), $26.6^\circ$ ($=\arctan(\frac{1}{2})$, $2$ edges), $45^\circ$ ($5$ edges), $63.4^\circ$ ($=\arctan(2)$, $2$ edges), $90^\circ$ ($5$ edges), and $-45^\circ$ ($2$ edges).}
\label{fig:ex1}
\end{figure}

\begin{example}{\rm
Let $\prm$ be the $2\times 3$ integer matrix
\[
\prm=\left[ \begin{matrix} 1 &0 &-1\\ 0 &1 &-1\end{matrix}\right].
\]
Then $M_\prm$ is the three-direction interpolating linear box spline with the refinement filter given by
\[
\wh{a_\prm}(\xi_1, \xi_2)=2^{-3} (1+e^{-i\xi_1})(1+e^{-i\xi_2})(1+e^{i(\xi_1+\xi_2)}).
\]
Note that $\mbox{supp}(a_\prm)=\{-1,0,1\}^2\bs\{(-1,1)^\tp,(1,-1)^\tp\}$ with $\#\mbox{supp}(a_\prm)=7$ and $\prm^{-1}(0,0)^\tp=\{(0,0,0)^\tp, (1,1,1)^\tp\}$, while $\prm^{-1} \gamma$ contains only one point in $\Z^3$ for every $\gamma\in \mbox{supp}(a_\prm)\bs\{(0,0)^\tp\}$.
Consequently, there are totally $21$ (by $\binom{7}{2}=21$) high-pass filters given by
\[
\frac{\sqrt{2}}{{8}}(\td_{(1,0)}-\td),\quad
\frac{\sqrt{2}}{{8}}(\td_{(1,1)}-\td),\quad
\frac{\sqrt{2}}{{8}}(\td_{(0,1)}-\td),\quad
\frac{\sqrt{2}}{{8}}(\td_{(-1,0)}-\td),\quad
\frac{\sqrt{2}}{{8}}(\td_{(-1,-1)}-\td),\quad
\frac{\sqrt{2}}{{8}}(\td_{(0,-1)}-\td)
\]
and all the rest $15$ filters are given by choosing a pair of two distinct points from $\mbox{supp}(a_\prm)\bs\{(0,0)^\tp\}$ (i.e., $\{ (1,0)^\tp, (1,1)^\tp, (0,1)^\tp, (-1,0)^\tp, (-1,-1)^\tp, (0,-1)^\tp\}$) with value $1/8$ at one point and $-1/8$ at the other point. This tight framelet filter bank $\{a_\prm; b_1,\ldots, b_{21}\}$ exhibits totally $6$ directions in dimension two. See Figure~\ref{fig:ex1} for details.
}\end{example}

\begin{figure}[ht]
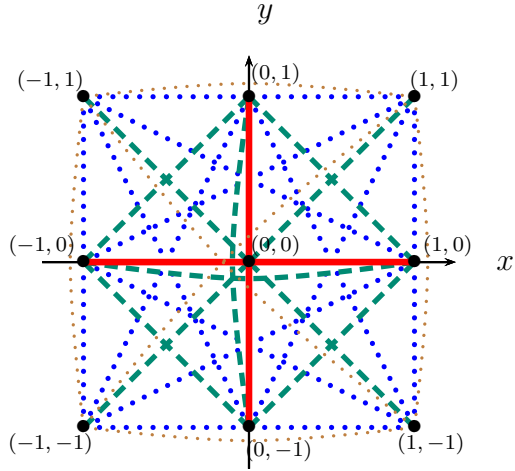

\psset{yunit=2.2cm,xunit=2.2cm}
\pspicture[](-1.25,-1.25)(1.25,1.25)
\psaxes[labels=none, ticks=none]{->}(0,0)(-1.25,-1.25)(1.25,1.25)
\rput[b](1.55,-0.05){$x$}
\rput[l](0.05,1.5){$y$}
\psline[linecolor=blue,linewidth=2pt,linestyle=dotted](1,-1)(1,0)
\psline[linecolor=blue,linewidth=2pt,linestyle=dotted](1,-1)(0,1)
\psline[linecolor=blue,linewidth=2pt,linestyle=dotted](1,-1)(-1,0)
\psline[linecolor=blue,linewidth=2pt,linestyle=dotted](1,-1)(0,-1)
\psline[linecolor=blue,linewidth=2pt,linestyle=dotted](-1,1)(1,0)
\psline[linecolor=blue,linewidth=2pt,linestyle=dotted](-1,1)(0,1)
\psline[linecolor=blue,linewidth=2pt,linestyle=dotted](-1,1)(-1,0)
\psline[linecolor=blue,linewidth=2pt,linestyle=dotted](-1,1)(0,-1)
\psline[linecolor=blue,linewidth=2pt,linestyle=dotted](1,1)(1,0)
\psline[linecolor=blue,linewidth=2pt,linestyle=dotted](1,1)(0,1)
\psline[linecolor=blue,linewidth=2pt,linestyle=dotted](1,1)(-1,0)
\psline[linecolor=blue,linewidth=2pt,linestyle=dotted](1,1)(0,-1)
\psline[linecolor=blue,linewidth=2pt,linestyle=dotted](-1,-1)(1,0)
\psline[linecolor=blue,linewidth=2pt,linestyle=dotted](-1,-1)(0,1)
\psline[linecolor=blue,linewidth=2pt,linestyle=dotted](-1,-1)(-1,0)
\psline[linecolor=blue,linewidth=2pt,linestyle=dotted](-1,-1)(0,-1)
\psline[linecolor=PineGreen,linewidth=2pt,linestyle=dashed](1,-1)(0,0)
\psline[linecolor=PineGreen,linewidth=2pt,linestyle=dashed](-1,1)(0,0)
\psline[linecolor=PineGreen,linewidth=2pt,linestyle=dashed](1,1)(0,0)
\psline[linecolor=PineGreen,linewidth=2pt,linestyle=dashed](-1,-1)(0,0)
\psline[linecolor=PineGreen,linewidth=2pt,linestyle=dashed](1,0)(0,1)
\psline[linecolor=PineGreen,linewidth=2pt,linestyle=dashed](1,0)(0,-1)
\psline[linecolor=PineGreen,linewidth=2pt,linestyle=dashed](0,1)(-1,0)
\psline[linecolor=PineGreen,linewidth=2pt,linestyle=dashed](-1,0)(0,-1)
\pscurve[linecolor=PineGreen,linewidth=2pt,linestyle=dashed](-1,0)(0,-0.1)(1,0)
\pscurve[linecolor=PineGreen,linewidth=2pt,linestyle=dashed](0,1)(-0.1,0)(0,-1)
\psline[linecolor=red,linewidth=2.5pt,linestyle=solid](1,0)(0,0)
\psline[linecolor=red,linewidth=2.5pt,linestyle=solid](0,1)(0,0)
\psline[linecolor=red,linewidth=2.5pt,linestyle=solid](-1,0)(0,0)
\psline[linecolor=red,linewidth=2.5pt,linestyle=solid](0,-1)(0,0)
\pscurve[linecolor=brown,linewidth=1.3pt,linestyle=dotted](1,-1)(-0.08,-0.08)(-1,1)
\pscurve[linecolor=brown,linewidth=1.3pt,linestyle=dotted](1,-1)(1.08,0)(1,1)
\pscurve[linecolor=brown,linewidth=1.3pt,linestyle=dotted](1,-1)(0,-1.08)(-1,-1)
\pscurve[linecolor=brown,linewidth=1.3pt,linestyle=dotted](-1,1)(0,1.08)(1,1)
\pscurve[linecolor=brown,linewidth=1.3pt,linestyle=dotted](-1,1)(-1.08,0)(-1,-1)
\pscurve[linecolor=brown,linewidth=1.3pt,linestyle=dotted](-1,-1)(-0.08,0.08)(1,1)
\rput(0,0){\black $\bullet$}\rput(0.16,0.1){\tiny $(0,0)$} 
\rput(1,0){\black $\bullet$}\rput(1.2,0.1){\tiny $(1,0)$} 
\rput(0,1){\black $\bullet$}\rput(0.16,1.14){\tiny $(0,1)$} 
\rput(-1,0){\black $\bullet$}\rput(-1.25,0.1){\tiny $(-1,0)$} 
\rput(0,-1){\black $\bullet$}\rput(0.18,-1.14){\tiny $(0,-1)$}
\rput(1,-1){\black $\bullet$}\rput(1.1,-1.1){\tiny $(1,-1)$} 
\rput(-1,-1){\black $\bullet$}\rput(-1.2,-1.1){\tiny $(-1,-1)$}
\rput(1,1){\black $\bullet$}\rput(1.12,1.1){\tiny $(1,1)$}  
\rput(-1,1){\black $\bullet$}\rput(-1.2,1.1){\tiny $(-1,1)$} 
\endpspicture
\caption{
Each edge connecting every two vertices indicates a high-pass filter with coefficients of the same weight but opposite signs at its two endpoints.
Weight for each small-dotted brown arc-edge (total 6) is $\frac{1}{16}$, weight for each dotted blue edge (total $16$) is $\frac{\sqrt{2}}{16}$, weight for each dashed green edge (total $10$) is $\frac{1}{8}$, and weight for each solid red edge (total $4$) is $\frac{\sqrt{2}}{8}$. The total number of directions/slopes for the $36$ high-pass filters is $8$ with angles $0^\circ$ ($9$ edges), $\pm26.6^\circ$ ($= \arctan(\frac{1}{2})$, $2$ edges each), $\pm 45^\circ$ ($5$ edges each), $\pm63.4^\circ$ ($=\arctan(\frac{1}{2})$, $2$ edges each), and $90^\circ$ ($9$ edges).}
\label{fig:ex2}
\end{figure}

\begin{example}\label{ex:tpl} {\rm
Let $\prm$ be the $2\times 4$ integer matrix
\[
\prm=\left[ \begin{matrix} 1 &0 &-1 &0\\ 0 &1 &0 &-1\end{matrix}\right].
\]
Then $M_\prm$ is the tensor product of the piecewise linear B-spline with the refinement filter $a_\prm$ given by
$\wh{a_\prm}(\xi_1, \xi_2)=2^{-2} |1+e^{-i\xi_1}|^2|1+e^{-i\xi_2}|^2$.
Note that $\mbox{supp}(a_\prm)=\{-1,0,1\}^2$ with $\#\mbox{supp}(a_\prm)=9$ and
\[
\begin{aligned}
\prm^{-1}(0,0)^\tp&=\{ (0,0,0,0)^\tp, (1,0,1,0)^\tp, (0,1,0,1)^\tp, (1,1,1,1)^\tp\},\\
\prm^{-1}(1,0)^\tp&=\{ (1,0,0,0)^\tp, (1,1,0,1)^\tp\},\quad
\prm^{-1}(0,1)^\tp=\{ (1,0,0,0)^\tp, (1,1,0,1)^\tp\},\\
\prm^{-1}(-1,0)^\tp&=\{ (0,0,1,0)^\tp, (0,1,1,1)^\tp\},\quad
\prm^{-1}(0,-1)^\tp=\{ (0,0,0,1)^\tp, (1,0,1,1)^\tp\},\\
\end{aligned}
\]
and $\prm^{-1} \gamma$ contains only one point in $\Z^4$ for every $\gamma\in
\{(1,-1)^\tp,(-1,1)^\tp,(1,1)^\tp,(-1,-1)^\tp\}$. Let
$S_1:=\{(1,-1)^\tp,(-1,1)^\tp,(1,1)^\tp,(-1,-1)^\tp\}$ and $S_2:=\{(1,0)^\tp,(0,1)^\tp,(-1,0)^\tp,(0,-1)^\tp\}$.
Consequently, there are total $36$ (by $\binom{9}{2}=36$) high-pass filters given by
\begin{align*}
&\frac{1}{{8}}(\td_{\gamma}-\td)\quad\forall\, \gamma\in S_1,
\qquad
\frac{\sqrt{2}}{{8}}(\td_{\gamma}-\td)\quad\forall\, \gamma\in S_2,
\qquad
\frac{\sqrt{2}}{{16}}(\td_{\gamma_1}-\td_{\gamma_2})\quad \forall\, \gamma_1\in S_1, \gamma_2\in S_2,\\
&\frac{1}{{16}}(\td_{\gamma_1}-\td_{\gamma_2})\quad\forall\, \gamma_1\neq\gamma_2, \gamma_1,\gamma_2\in S_1,\qquad
\frac{1}{{8}}(\td_{\gamma_1}-\td_{\gamma_2})\quad \forall\, \gamma_1\neq\gamma_2, \gamma_1,\gamma_2\in S_2
\end{align*}
with $4$, $4$, $16$, $6$, and $6$ filters for each group, respectively.
This tight framelet filter bank $\{a_\prm; b_1,\ldots, b_{36}\}$ exhibits totally $8$ directions in dimension two. See Figure~\ref{fig:ex2} for details.
}\end{example}

Note that if $\{a;b_1,\ldots,b_s\}$ is a tight framelet filter bank, then $\{a; b_1(\cdot-2k_1),\ldots,b_s(\cdot-2k_s)\}$ is also a tight framelet filter bank for all $k_1,\ldots,k_s\in \dZ$. Using this simple observation, the number of the high-pass filters in Example~\ref{ex:tpl} can be reduced from $36$ to $30$. Indeed,
the two filters corresponding to the edges $\{(1,-1)^\tp, (1,1)^\tp\}$ and $\{(-1,-1)^\tp, (-1,1)^\tp\}$ can be combined into one single filter; the two filters for the edges $\{(0,1)^\tp,(1,1)^\tp\}$ and $\{(0,-1)^\tp,(1,-1)^\tp\}$ can be combined into one filter; the two filters for the edges $\{(-1,1)^\tp,(0,1)^\tp\}$ and $\{(-1,-1)^\tp,(0,-1)^\tp\}$ can be combined into one filter. We can also perform the same operation for filters with the vertical direction. Using this and more complicated argument/technique, the number of high-pass filters in a box spline tight framelet filter bank in Theorem~\ref{thm:boxspline} often can be reduced. We shall not address this issue in this paper.
We shall explore some applications of the directional tight framelets in Theorems~\ref{thm:haar} and~\ref{thm:boxspline} elsewhere.


\begin{thebibliography}{10}
\bibitem{bhr93}
C. de Boor, K. H\"ollig, S. Riemenschneider. Box splines.
Series in Appl. Math. Sci. \textbf{98}. Springer-Verlag, New York, 1993.

\bibitem{CD04}
E.~J.~Cand\`es and D.~L.~Donoho, New tight frames of curvelets and optimal representations of objects with piecewise $C^2$ singularities. \emph{Comm. Pure Appl. Math.} \textbf{57} (2004), 219--266.

\bibitem{chs02}
C.~K.~Chui, W.~He, and J.~St\"ockler, Compactly supported tight and sibling frames with maximum vanishing moments. \emph{Appl. Comput. Harmon. Anal.} \textbf{13} (2002), 224--262.

\bibitem{dgm86}
I.~Daubechies, A.~Grossmann, and Y.~Meyer,
Painless nonorthogonal expansions.
\emph{J. Math. Phys.} \textbf{27} (1986), 1271--1283.

\bibitem{dhrs03}
I.~Daubechies, B.~Han, A.~Ron, Z.~Shen, Framelets: MRA-based constructions of wavelet frames. \emph{Appl. Comput. Harmon. Anal.} \textbf{14} (2003), 1--46.

\bibitem{fjs16}
Z.~Fan, H.~Ji, and Z.~Shen, Dual Gramian analysis: duality principle and unitary extension principle. \emph{Math. Comp.} \textbf{85} (2016), 239--270.

\bibitem{GKL06}
K. Guo, G. Kutyniok, and D. Labate, {Sparse multidimensional
representations using anisotropic dilation and shear operators}, Wavelets and Splines (Athens, GA, 2005), Nashboro Press, Nashville,
TN (2006) 189-201.

\bibitem{han14}
B.~Han, The projection method for multidimensional framelet and wavelet analysis. \emph{Math. Model. Nat. Phenom.} \textbf{9} (2014), 83--110.

\bibitem{han13}
B.~Han, Properties of discrete framelet transforms. \emph{Math. Model. Nat. Phenom.} \textbf{8} (2013), 18--47.

\bibitem{han12}
B.~Han, Nonhomogeneous wavelet systems in high dimensions. \emph{Appl. Comput. Harmon. Anal.} \textbf{32} (2012), 169--196.

\bibitem{han06}
B. Han, The porjection method in wavelet analysis, in Splines and Wavelets: Athens 2005, G. Chen and M.J. Lai eds., (2006), 202--225.

\bibitem{han03}
B.~Han, Compactly supported tight wavelet frames and orthonormal wavelets of exponential decay with a general dilation matrix. \emph{J. Comput. Appl. Math.} \textbf{155} (2003), 43--67.

\bibitem{han99}
B.~Han, Analysis and construction of optimal multivariate biorthogonal wavelets with compact support. \emph{SIAM J. Math. Anal.} \textbf{31} (1999), 274--304.

\bibitem{han97}
B.~Han, On dual wavelet tight frames. \emph{Appl. Comput. Harmon. Anal.} \textbf{4} (1997), 380--413.

\bibitem{hz14}
B.~Han and Z.~Zhao, Tensor product complex tight framelets with increasing directionality. \emph{SIAM J. Imaging Sci.} \textbf{7} (2014), 997--1034.

\bibitem{ls06}
M.~J.~Lai and J.~St\"ockler, Construction of multivariate compactly supported tight wavelet frames. \emph{Appl. Comput. Harmon. Anal.} \textbf{21} (2006), 324--348.

\bibitem{lcshi16}
Y.-R.~Li, R.~H.~Chan, L.~Shen, Y.-C.~Hsu, and W.-Y. Tseng, An adaptive directional Haar framelet-based reconstruction algorithm for parallel magnetic resonance imaging. \emph{SIAM J. Imaging Sci.} \textbf{9} (2016), 794--821.

\bibitem{rs97}
A.~Ron and Z.~Shen, Affine systems in $L_2(\R^d)$: the analysis of the analysis operator. \emph{J. Funct. Anal.} \textbf{148} (1997), 408--447.

\bibitem{rs98}
A.~Ron and Z.~Shen, Compactly supported tight affine spline frames in $L_2(\R^d)$. \emph{Math. Comp.} \textbf{67} (1998), 191--207.



\end{thebibliography}
\end{document}